\documentclass[a4paper,preprint,amsmath,amssymb,12pt,superbib]{revtex4}

\usepackage{epsfig,graphicx,dcolumn,bm,times}

\begin{document}

\title{Projective-anticipating, projective, and projective-lag synchronization of time-delayed chaotic systems on random networks}

\author{Cun-Fang Feng,$^{1}$ Xin-Jian Xu,$^{2,3}$ Sheng-Jun Wang,$^{1}$ and Ying-Hai Wang$^{1,}$\footnote{For correspondence: yhwang@lzu.edu.cn}}

\address{$^{1}$Institute of Theoretical Physics, Lanzhou University, Lanzhou Gansu 730000, China\\
$^{2}$Departamento de F\'{i}sica da Universidade de Aveiro,
3810-193 Aveiro, Portugal\\$^{3}$Department of Mathematics,
College of Science, Shanghai University, Shanghai 200444, China}

\date\today

\begin{abstract}
We study projective-anticipating, projective, and projective-lag
synchronization of time-delayed chaotic systems on random
networks. We relax some limitations of previous work, where
projective-anticipating and projective-lag synchronization can be
achieved only on two coupled chaotic systems. In this paper, we
can realize projective-anticipating and projective-lag
synchronization on complex dynamical networks composed by a large
number of interconnected components. At the same time, although
previous work studied projective synchronization on complex
dynamical networks, the dynamics of the nodes are coupled
partially linear chaotic systems. In this paper, the dynamics of
the nodes of the complex networks are time-delayed chaotic systems
without the limitation of the partial-linearity. Based on the
Lyapunov stability theory, we suggest a generic method to achieve
the projective-anticipating, projective, and projective-lag
synchronization of time-delayed chaotic systems on random
dynamical networks and find both the existence and sufficient
stability conditions. The validity of the proposed method is
demonstrated and verified by examining specific examples using
Ikeda and Mackey-Glass systems on Erd\"{o}s-R\'{e}nyi networks.
\\
\\
\emph{PACS}: 05.45.Xt, 05.45.Jn, 89.75.Hc
\\
\\
\emph{Keywords}: projective-anticipating synchronization; projective synchronization;
projective-lag synchronization; time-delayed chaotic systems; complex networks
\end{abstract}

\maketitle

\bfseries

In 1999, Mainieri and Rehacek observed projective synchronization
in coupled partially linear chaotic systems where the drive and
response vectors synchronize up to a constant ratio $\alpha$
(scaling factor). Complete synchronization and anti-phase
synchronization are proved to be the special cases of projective
synchronization in cases of $\alpha=1$ and $\alpha=-1$,
respectively. This proportional feature can be used to extend
binary digital to M-nary digital for achieving fast communication.
With the development of research on complex systems, more and more
researchers carried out the study about complex dynamical
behaviors on networks. While most studies focused on complete and
phase synchronization in various networks, little attention has
been paid to projective synchronization. In this paper, based on
the Lyapunov stability theory, we theoretically analyze both the
existence and sufficient stability conditions of the
projective-anticipating, projective, and projective-lag
synchronization of time-delayed chaotic systems on random
networks. Numerical experiments for the Ikeda system and the
Mackey-Glass system show that the control method works.

\mdseries

\section{INTRODUCTION}

Chaos synchronization has attracted considerably increasing
attention and become an active area of research for both
theoretical interests and practical applications, since seminal
work of Pecora and Carroll \cite{Pecora}. Over the last decade,
following the complete synchronization \cite{Pecora}, several new
types of synchronization have been found in interacting chaotic
systems, such as  generalized synchronization \cite{Rulkov}, phase
synchronization \cite{Rosenblum}, anti-phase synchronization
\cite{Cao}, projective synchronization \cite{Mainieri}, lag
synchronization \cite{M.G. Rosenblum} and anticipating
synchronization \cite{H.U. Voss}. Complete synchronization is
characterized by the convergence of the two chaotic trajectories,
$y(t)=x(t)$. It appears only when interacting systems are
identical. Generalized synchronization means the amplitude of the
slave's state variable correlated with that of master's by a
generic function, $y(t)=F(x(t))$. Phase synchronization is defined
as the entrainment of phases of chaotic oscillators,
$n\Phi_{x}-m\Phi_{y}=\texttt{constant}$ ($n$ and $m$ are
integers), whereas their amplitude remains chaotic and
uncorrelated. Projective synchronization is the dynamical behavior
in which the amplitude of the master's state variable and that of
the slave's synchronizes up to a constant scaling factor $\alpha$
(a proportional relation). Complete synchronization and anti-phase
synchronization are the special cases of the projective
synchronization in cases of $\alpha=1$ and $\alpha=-1$,
respectively. Lag synchronization means a coincidence of
shifted-in-time states of two coupled systems, the state variable
of the drive system is delayed by positive $\tau$ in comparison
with that of the driven: $y(t)=x(t-\tau)$, $(\tau>0)$. Whereas for
anticipating synchronization, the driven system anticipates the
driver: $y(t)=x(t+\tau)$, $(\tau>0)$. Among those synchronization,
projective synchronization is one of the most interesting
problems, because of its proportion between the synchronized
dynamical states. In application to secure communications, this
feature can be used to extend binary digital to M-nary digital
communication \cite{Chee} for achieving fast communication. The
early study of projective synchronization reported that the
projective synchronization was usually observable only in the
coupled partially linear systems \cite{Xu}. Following work has
extended that to a general class of chaotic systems without the
limitation of partial-linearity \cite{Wen,feng}. Recently, Hoang
\emph{et al}. investigated a new synchronization in time-delay
chaotic system, which they called projective-anticipating
synchronization \cite{Hoang1}. Projective-anticipating
synchronization is a combination of the well-known schemes of
projective and anticipating synchronization. That is, the driver
synchronizes with the driven under the anticipating
synchronization scheme and the amplitude is correlated by a
scaling factor $\alpha$: $y(t)=\alpha x(t+\tau), (\tau>0)$. In the
case of projective-lag synchronization \cite{Hoang}, the amplitude
of the master's and slave's state variables is correlated by a
scale factor $\alpha$: $y(t)=\alpha x(t-\tau), (\tau>0)$.

However, most existing work about projective-anticipating and
projective-lag synchronization focused only on two coupled chaotic
systems. Systems consisting of many interconnected subsystems are
ubiquitous in nature and social science which can be described by
complex networks \cite{Albert}. Complex networks are usually
composed by a large number of interconnected components (nodes)
representing individuals or organizations and edges mimicking the
interaction among them. With the development of research on
complex systems, more and more researchers carried out the study
about complex dynamical behaviors on networks. While most studies
focused on complete \cite{Nishikawa} and phase synchronization
\cite{Hong} in various networks, little attention has been paid to
projective synchronization. Very recently, Hu et al. \cite{Hu}
studied the projective synchronization on drive-response dynamical
networks by considering coupled Lorenz chaotic systems. But due to
finite signal transmission times, switching speeds and memory
effects with both single and multiple delays are ubiquitous in
nature, technology and society \cite{Traub,Foss}. Time-delayed
systems are also interesting because the dimension of their
chaotic dynamics can be increased by increasing the delay time
sufficiently \cite{Pyragas}. From this point of view, these
systems are especially appealing for secure communication schemes.
In addition, time-delayed system can be considered as a special
case of spatiotemporal systems \cite{Masoller}. So it is natural
to consider an interesting topic whether we can achieve
projective-anticipating, projective, and projective-lag
synchronization of time-delayed chaotic systems on complex
dynamical networks? In this paper, inspired by the above
discussions, we extends the work on projective-anticipating,
projective, and projective-lag synchronization of two coupled
chaotic systems to complex dynamical networks. We attempt to
achieve projective-anticipating, projective and projective-lag
synchronization in a general class of time-delayed chaotic systems
related to optical bistable or hybrid optical bistable device on
complex dynamical networks.

The layout of this paper is as follows. In Sec. II, according to
the Lyapunov stability theory \cite{R.He}, we theoretically
analyze both the existence and sufficient stability conditions of
the projective-anticipating, projective, and projective-lag
synchronization of time-delayed chaotic systems on complex
dynamical networks. In Sec. III, the well-known Ikeda and
Mackey-Glass systems are considered as the dynamics of single node
to prove the validity of the proposed theoretical approach in Sec.
II, respectively. Finally we end this paper by a short conclusion
in last section.

\section{THEORETICAL ANALYSIS USING A GENERIC MODEL}

For convenience, we study different types of projective
synchronization of infinite-dimensional chaotic systems on
Erd\"{o}s-R\'{e}nyi (ER) random networks \cite{Erdos}. Consider an
ER network consisting of $1+N$ nodes which constitute the
drive-response dynamical networks \cite{Hu}. In response dynamical
networks, we connect each pair of nodes with probability $p$, and
denote the degree of node $i$ with $k_{i}$. Each individual node
is a delay-differential system related to the optical bistable or
hybrid optical device \cite{Li}
\begin{equation}
\tau^{'}\dot{x}(t)=-\beta x(t)+\mu f(x(t-t_{R})) \label{single1}
\end{equation}
where $x(t)$ is the dimensionless output of the system at time
$t$, $t_{R}$ is the delay time of the feedback loop, $\tau^{'}$ is
the response time of the nonlinear medium, $\mu$ is proportional
to the intensity of the incident light and $\beta$ is the
parameter. In Eq.(\ref{single1}), $f(x)$ is a nonlinear function
of $x$, characterizing the system, e.g.,
$f(x)=\pi[1+2B\cos(x+x_{0})]$ for Ikeda model \cite{Ikeda},
$f(x)=\pi[A-\sin^{2}(x-x_{0})]$ for Vall\'{e}e model
\cite{Vallee}, $f(x)=\sin^{2}(x-x_{0})$ for the sine-square model
\cite{Goedgebuer}, and $f(x)=ax/(1+x^{c})$ for Mackey-Glass model
\cite{Mackey}. In order to observe the different types of
projective synchronization, the drive-response dynamical networks
are described by the following equations:
\begin{subequations}
\begin{equation}
\tau^{'}\dot{x}(t)=-\beta x(t)+\mu_{1}f(x(t-\tau_{1})),
\label{driver1}
\end{equation}
{\setlength\arraycolsep{2pt}
\begin{eqnarray}
\tau^{'}\dot{y_{i}}(t)&=&-\beta
y_{i}(t)+\mu_{2}f(\frac{y_{i}(t-\tau_{1})}{\alpha})+\mu
f(x(t-\tau_{2})) +\epsilon \sum_{j=1}^N G_{ij}y_{j},
(i=1,2,\ldots,N)\label{responser1}
\end{eqnarray}}
\end{subequations}
where $x$ denotes the drive system, $y_{i}$ denotes the response
dynamical network systems, $\epsilon$ is the coupling strength,
$\alpha$ is a desired scaling factor present for the
projective-anticipating, projective, and projective-lag
synchronization, $\tau_{1}$ is the feedback delay time in the
coupled systems and $\tau_{2}$ is the coupling delay time between
systems $x$ and $y_{i}$. The graph topology is encoded in the
Laplacian G, a symmetric matrix with zero row-sum,
\begin{equation}
G_{ij}=\left\{ \begin{array}{ll}
 -k_{i},& \textrm{if $i=j$}, \\
 1, & \textrm{if node $i$ connects node $j$},\\
 0, & \textrm{otherwise}.
 \end{array} \right.
\label{Laplacian}
\end{equation}
The eigenvalues $\lambda_{i}$ of $G$ are real and nonpositive
\cite{Jost}.

Throughout this study, for simplification we use the notation
$x_{\tau}\equiv x(t-\tau)$. In the following, we will investigate
different kinds of projective synchronization. One can find that
under conditions
\begin{equation}
\mu=\alpha \mu_{1}-\mu_{2}, \label{existence}
\end{equation}
Eqs. (\ref{driver1}) and (\ref{responser1}) allow for
synchronization manifold
\begin{equation}
y_{i}=\alpha x_{\tau_{2}-\tau_{1}}. \label{manifold}
\end{equation}
We define the error systems as the difference between the Eqs.
(\ref{driver1}) and (\ref{responser1}): $e_{i}=y_{i}-\alpha
x_{\tau_{2}-\tau_{1}}$, one obtains error dynamics
\begin{equation}
\begin{split}
\dot{e}_{i}&=\dot{y}_{i}-\alpha\dot{x}_{\tau_{2}-\tau_{1}}\\
           &=\frac{1}{\tau^{'}}[-\beta
           e_{i}+\mu_{2}f(\frac{y_{i,\tau_{1}}}{\alpha})+\mu f(x_{\tau_{2}})-\alpha \mu_{1} f(x_{\tau_{2}})+\epsilon\sum_{j=1}^N
           G_{ij}e_{j}],\\
\label{error}
\end{split}
\end{equation}
according to Eq.(\ref{manifold}),
$f(\frac{y_{i,\tau_{1}}}{\alpha})$ contained in Eq.(\ref{error})
can be expressed as $f(x_{\tau_{2}})$. Therefore, combining
Eq.(\ref{existence}), Eq.(\ref{error}) can be reduced as
\begin{equation}
\dot{e}_{i}=\frac{1}{\tau^{'}}[-\beta
           e_{i}+\epsilon\sum_{j=1}^N
           G_{ij}e_{j}].
\label{error2}
\end{equation}
Consider a Lyapunov function in the form
\begin{equation}
V=\frac{1}{2} \sum_{i=1}^N e_{i}^{2},\label{Lyapunov}
\end{equation}
according to the Lyapunov stability theory \cite{R.He}, if the
function (\ref{Lyapunov}) satisfies the following conditions
\begin{equation}
\left\{ \begin{array} {ll} V(\textbf{e})>0
&\textrm{if $\textbf{e} \neq 0$,}\\
\\
V(\textbf{e})=0 &\textrm{if $\textbf{e}=0$,}
\end{array}\right.
\label{stability1}
\end{equation}
and
\begin{equation}
\left\{ \begin{array} {ll} \dot{V}(\textbf{e})<0
&\textrm{if $\textbf{e} \neq 0$,}\\
\\
\dot{V}(\textbf{e})=0 &\textrm{if $\textbf{e}=0$,}
\end{array}\right.
\label{stability2}
\end{equation}
then $\textbf{e}$ will asymptotically converge to zero as time
tends to infinity leading to
$\lim_{t\rightarrow\infty}\|y_{i}-\alpha
x_{\tau_{2}-\tau_{1}}\|=0$, i.e., projective-anticipating,
projective, and projective-lag synchronization of time-delayed
chaotic systems can be realized on ER networks.

Surely, the Lyapunov function (\ref{Lyapunov}) satisfies the
condition (\ref{stability1}). For the condition
(\ref{stability2}), from Eqs. (\ref{error2}) and (\ref{Lyapunov}),
we get
\begin{equation}
\begin{split}
\dot{V}&=\sum_{i=1}^N e_{i}\dot{e_{i}}\\
       &=\frac{-\beta}{\tau^{'}}\sum_{i=1}^N e_{i}^{2}+\frac{\epsilon}{\tau^{'}}\sum_{i=1}^N \sum_{j=1}^N
          e_{i}G_{ij}e_{j}\\
       &=\textbf{e}G^{'}\textbf{e}^{T},\\
\label{stability3}
\end{split}
\end{equation}
where
$G^{'}=\frac{\epsilon}{\tau^{'}}G-\frac{\beta}{\tau^{'}}I_{N}$,
$\textbf{e}=(e_{1}, e_{2}, \ldots,\ldots,e_{N})$. Because $G$ have
zero row-sum matrix, according to the stability theory and the
Gerschgorin's disk theorem, we can obtain that the sufficient
stability condition for the synchronized manifold (\ref{manifold})
can be written as $\frac{\beta}{\tau^{'}}>0$ and
$\frac{\epsilon}{\tau^{'}}>0$. Under this sufficient stability and
existence conditions (\ref{existence}), with the change of
magnitude relation between feedback delay time $\tau_{1}$ and
coupling delay time $\tau_{2}$, we can observe different types of
projective synchronization. Let
$\Delta_{\tau}=|\tau_{2}-\tau_{1}|$. If $\tau_{2}<\tau_{1}$, the
synchronization manifold $y_{i}=\alpha x_{\tau_{2}-\tau_{1}}$
(\ref{manifold}) can be written as $y_{i}(t)=\alpha
x(t+\Delta_{\tau})$, i.e., we can observe projective-anticipating
synchronization. If $\tau_{2}=\tau_{1}$, the synchronization
manifold $y_{i}=\alpha x_{\tau_{2}-\tau_{1}}$ (\ref{manifold}) can
be written as $y_{i}(t)=\alpha x(t)$, i.e., we can observe exact
projective synchronization. When $\tau_{2}>\tau_{1}$, we can
derive $y_{i}(t)=\alpha x(t-\Delta_{\tau})$, that is to say,
projective-lag synchronization can be observed.

\textbf{Remark 1} The value of scaling factor $\alpha$ has no
effect on the error dynamics of the system (Eq. (\ref{error2}))
because the values of $\dot{e_{i}}$ are independent of the scaling
factor $\alpha$. So we can arbitrarily direct the scaling factor
$\alpha$ onto any desired value.

\section{NUMERICAL SIMULATIONS}

In this section, we give two illustrative examples.

\textbf{Remark 2} Because the whole response dynamical networks
realize complete synchronization, we can get the
projective-anticipating, projective, and projective-lag
synchronization between the drive system and the response
dynamical networks with the same scaling factor $\alpha$.

\textbf{Remark 3} In the first example and the second one, the ER
networks consist of $1+N=1001$ nodes. In response dynamical
networks systems, connectivity probability is $p=0.01$. Because
the size of ER networks is large, it is difficult to get the
numerical results between the drive system and the whole response
dynamical networks. In simulations, we give the numerical results
with three arbitrarily selected response systems. If they can
achieve projective-anticipating, projective, and projective-lag
synchronization with the drive system $x$, the other response
systems can do it with the same scaling factor $\alpha$.

\textbf{Example 1.} Consider the nodes of the drive-response
dynamical networks are Ikeda systems as follows:
\begin{subequations}
\begin{equation}
\dot{x}(t)=-\beta x(t)+\mu_{1}sin(x_{\tau_{1}}), \label{driver3}
\end{equation}
{\setlength\arraycolsep{2pt}
\begin{eqnarray}
\dot{y_{i}}(t)&=&-\beta
y_{i}(t)+\mu_{2}sin(\frac{y_{i,\tau_{1}}}{\alpha})+\mu
sin(x_{\tau_{2}}) +\epsilon\sum_{j=1}^N G_{ij}y_{j},
(i=1,2,\ldots,N) \label{responser3}
\end{eqnarray}}
\end{subequations}
with $\beta>0$ and $\mu_{1,2}<0$. Here, $\beta$ is the relation
coefficient for the drive system $x$ and response dynamical
networks $y_{i}$, $\mu_{1,2}$ is proportional to the power of the
incident light, the delay $\tau_{1}$ is the time required for
light to make a round trip in the cavity. Ikeda model was
introduced to investigate the dynamics of an optical bistable
resonator, playing an important role in electronical and
physiological study and is well known for delay-induced chaotic
behavior \cite{H.U. Voss, E. M. Shahverdiev}.

\begin{figure*}
\includegraphics[width=18cm]{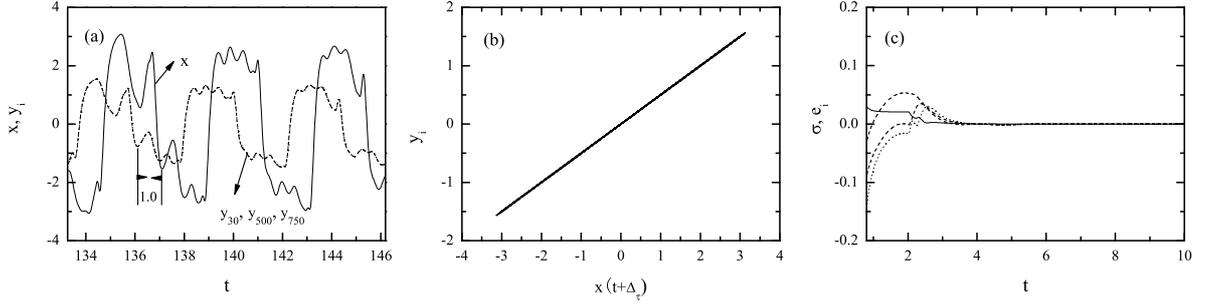}

\caption{Projective-anticipating synchronization for $\alpha=0.5,
\beta=5, \tau_{1}=2, \tau_{2}=1, \mu_{1}=-16, \mu_{2}=-0.2,
\mu=-7.8$: (a) The time series of the driver system $x(t)$ (solid
line) and the driven systems $y_{i}(t)$ (dotted line); (b)
synchronization manifold between $x(t+\Delta_{\tau})$ and $y(t)$,
$\Delta_{\tau}=|\tau_{2}-\tau_{1}|$; (c) the time series of the
error systems $\sigma$ (solid line) and $e_{i}$ (dotted
line).}\label{fig1}
\end{figure*}

\begin{figure*}
\includegraphics[width=18cm]{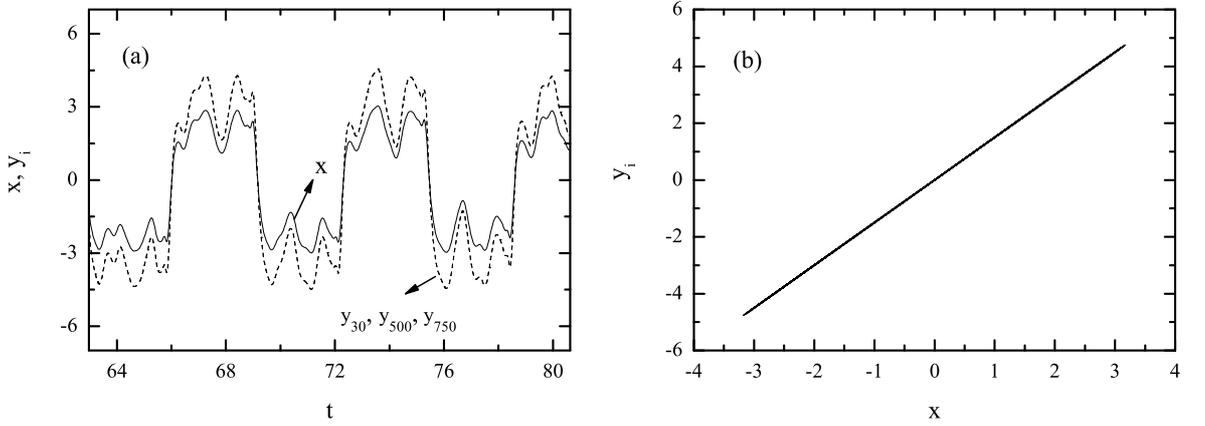}

\caption{Exact projective synchronization for $\alpha=1.5,
\beta=5, \tau_{1}=3, \tau_{2}=3, \mu_{1}=-16, \mu_{2}=-0.4,
\mu=-23.6$: (a) The time series of the driver system $x(t)$ (solid
line) and the driven systems $y_{i}(t)$ (dotted line), (b) the
synchronization between $x(t)$ and $y_{i}(t)$.}\label{fig2}
\end{figure*}

\begin{figure*}
\includegraphics[width=18cm]{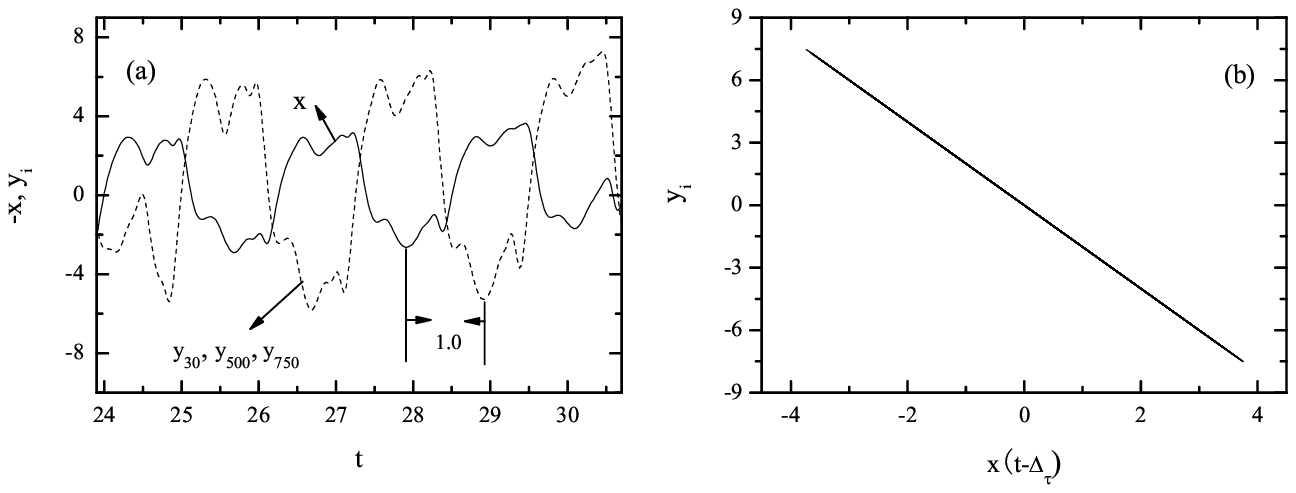}
\caption{Projective-lag synchronization for $\alpha=-2.0, \beta=5,
\tau_{1}=1, \tau_{2}=2, \mu_{1}=-20, \mu_{2}=-0.5, \mu=40.5$: (a)
The time series of the driver system $-x(t)$ (solid line) and the
driven systems $y_{i}(t)$ (dotted line); (b) the synchronization
between $x(t-\Delta_{\tau})$ and $y_{i}(t)$,
$\Delta_{\tau}=\tau_{2}-\tau_{1}$.}\label{fig3}
\end{figure*}

In the following, we will confirm that the numerical simulations
fully support the analytical results presented above. The coupling
strength in response dynamical networks is set as $\epsilon=0.1$.
If $\tau_{2}<\tau_{1}$, one can observe the
projective-anticipating synchronization for parameters
$\alpha=0.5, \beta=5, \tau_{1}=2, \tau_{2}=1, \mu_{1}=-16,
\mu_{2}=-0.2, \mu=-7.8$, with the anticipating time
$\Delta_{\tau}=|\tau_{2}-\tau_{1}|=1$ . In Fig. \ref{fig1}(a), we
can observe that the driven systems anticipates the driver, at the
same time, the amplitude of $x(t+1)$ and $y_{i}(t)$ correlates
with each other by $y_{i}(t)=0.5x(t+1)$. Figure \ref{fig1}(b)
shows the time-shifted plot of $x(t+1)$ and $y_{i}(t)$. It is
clear from the scales of the coordinate axes that the slope of the
line is $0.5$. In Fig. \ref{fig1}(c), we give numerical results of
the time series of the error systems
$\sigma=\frac{1}{N}\sum^{N}_{i=1}|y_{i}(t)-<y_{i}(t)>|$ and
$e_{i}$, where $<\cdot>$ indicates the average of nodes of
response dynamical networks. From it we can see that the whole
response dynamical networks already realize complete
synchronization when projective-anticipating synchronization among
drive-response complex dynamical networks can be observed. With
$\tau_{1}=\tau_{2}$, we can observe the exact projective
synchronization between Eqs. (\ref{driver3}) and
(\ref{responser3}) for parameters $\alpha=1.5, \beta=5,
\tau_{1}=3, \tau_{2}=3, \mu_{1}=-16, \mu_{2}=-0.4, \mu=-23.6$.
From Fig. \ref{fig2}(a), one can find that the phase angle between
the synchronized trajectories is zero. It reduces to the complete
synchronization if $\alpha=1.0$. From Fig. \ref{fig2}(b), we can
see the synchronization between $x(t)$ and $y_{i}(t)$ for
$\alpha=1.5$. For coupling delay $\tau_{2}$ being larger than
feedback delay $\tau_{1}$, we can observe the projective-lag
synchronization for parameters $\alpha=-2.0, \beta=5, \tau_{1}=1,
\tau_{2}=2, \mu_{1}=-20, \mu_{2}=-0.5, \mu=40.5$. In Fig.
\ref{fig3}(a), $-x(t)$ instead of $x(t)$ is convenient for our
study. The response systems lag the state of the drive system with
constant lag time $\Delta_{\tau}=\tau_{2}-\tau_{1}=1$ and the
amplitude of the driver's and driven's state variables is
correlated by $y_{i}(t)=-2x(t-1)$. The time-shifted plot of
$x(t-1)$ and $y(t)$ is shown in Fig. \ref{fig3}(b). It is also
clear from the scales of the coordinate axes that the slope of the
line is $-2.0$.

\begin{figure*}
\includegraphics[width=18cm]{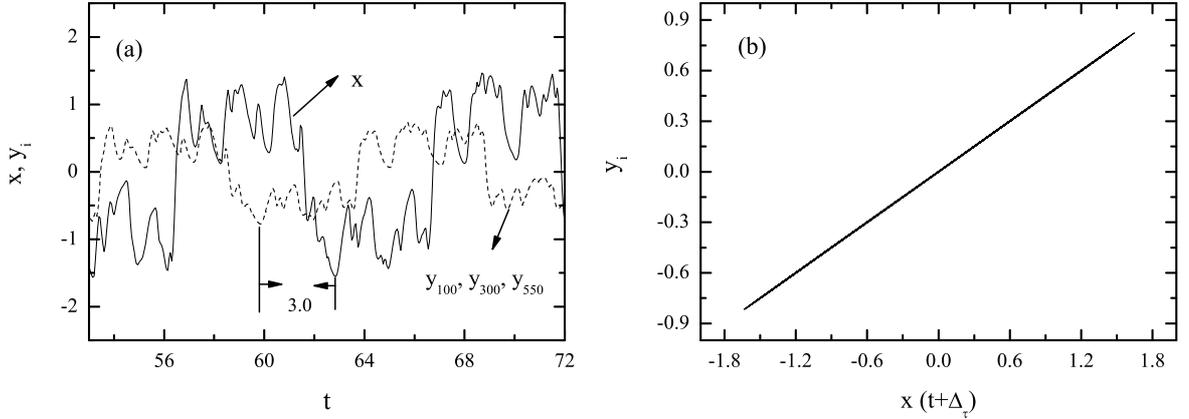}

\caption{Projective-anticipating synchronization for $\alpha=0.5,
\beta=6, \tau_{1}=5, \tau_{2}=2, \mu_{1}=-14, \mu_{2}=-0.2,
\mu=-6.8$: (a) The time series of the driver system $x(t)$ (solid
line) and the driven systems $y_{i}(t)$ (dotted line); (b)
synchronization manifold between $x(t+\Delta_{\tau})$ and $y(t)$,
$\Delta_{\tau}=|\tau_{2}-\tau_{1}|$.}\label{fig4}
\end{figure*}

\begin{figure*}
\includegraphics[width=18cm]{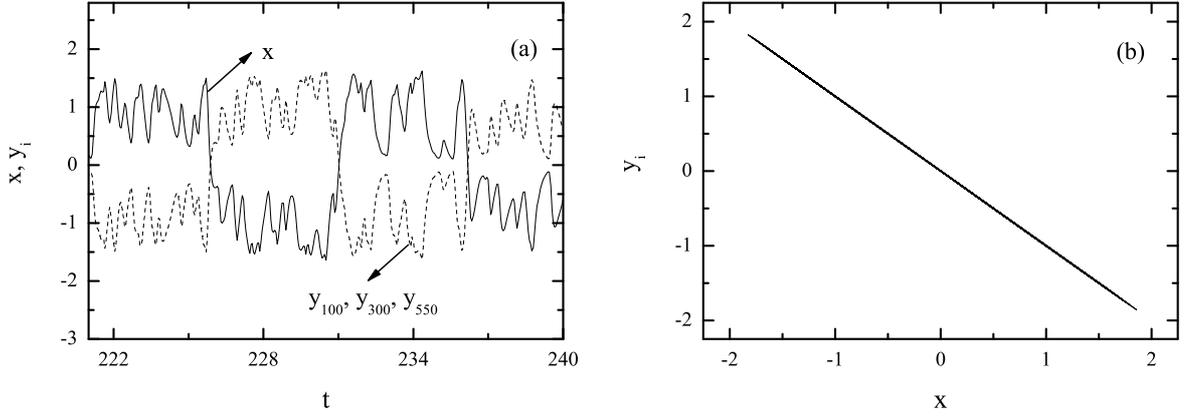}

\caption{Anti-phase synchronization for $\alpha=-1.0, \beta=6,
\tau_{1}=5, \tau_{2}=5, \mu_{1}=-16, \mu_{2}=-0.2, \mu=16.2$: (a)
The time series of the driver system $x(t)$ (solid line) and the
driven systems $y_{i}(t)$ (dotted line), (b) the synchronization
between $x(t)$ and $y_{i}(t)$.}\label{fig5}
\end{figure*}

\begin{figure*}
\includegraphics[width=18cm]{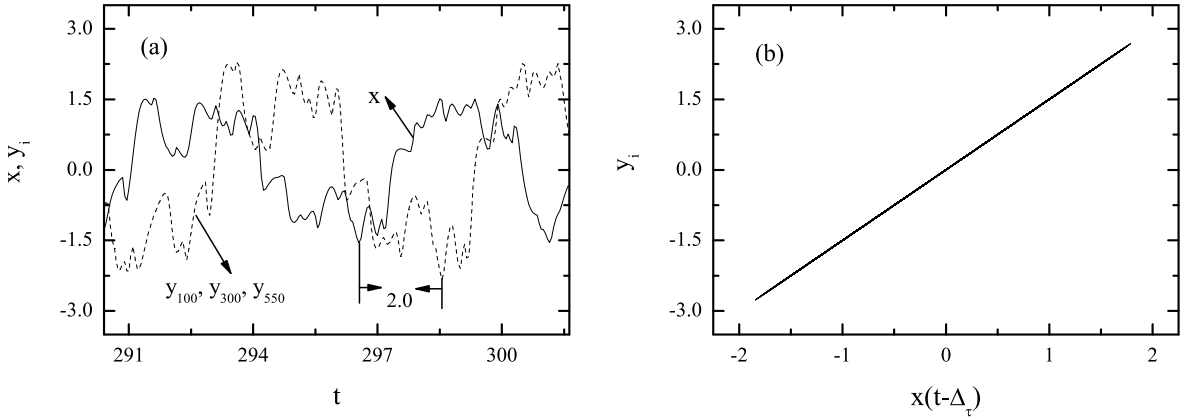}

\caption{Projective-lag synchronization for $\alpha=1.5, \beta=6,
\tau_{1}=3, \tau_{2}=5, \mu_{1}=-16, \mu_{2}=-0.2, \mu=-23.8$: (a)
The time series of the driver system $x(t)$ (solid line) and the
driven systems $y_{i}(t)$ (dotted line); (b) the synchronization
between $x(t-\Delta_{\tau})$ and $y_{i}(t)$,
$\Delta_{\tau}=\tau_{2}-\tau_{1}$.}\label{fig6}
\end{figure*}

\textbf{Example 2.} Consider the nodes of the drive-response
dynamical networks are Mackey-Glass systems as follows:
\begin{subequations}
\begin{equation}
\dot{x}(t)=-\beta x(t)+\mu_{1}x_{\tau_{1}}/(1+x^{b}_{\tau_{1}}).
\label{driver4}
\end{equation}
Initially, it was introduced as a model of blood generation for
patients with leukemia. $x(t)$ represents the density of
circulating cells at time $t$, when it is produced, and
$x_{\tau_{1}}$ is the density when the "request" for more blood
cells is made. Later, this system became popular in chaos theory
as a model for producing high-dimensional chaos to test various
methods of chaotic time-series analysis, controlling chaos, etc.
The electronic analog of this system has been proposed by Pyragas
and his collaborators \cite{A. Namajunas}. The response dynamical
network systems are written as {\setlength\arraycolsep{2pt}
\begin{eqnarray}
\dot{y_{i}}(t)=-\beta
y_{i}(t)+\mu_{2}(y_{i,\tau_{1}}/\alpha)/(1+(y_{i,\tau_{1}}/\alpha)^{b})+\mu
x_{\tau_{2}}/(1+x^{b}_{\tau_{2}})+\epsilon\sum_{j=1}^N
G_{ij}y_{j},
                 \nonumber\\
(i=1,2,\ldots,N). \label{responser4}
\end{eqnarray}}
\end{subequations}

In our simulation, the coupling strength in response dynamical
networks is set as $\epsilon=0.2$. With $\tau_{2}$ being less than
$\tau_{1}$, one can observe the projective-anticipating
synchronization for parameters $\alpha=0.5, \beta=6, \tau_{1}=5,
\tau_{2}=2, \mu_{1}=-14, \mu_{2}=-0.2, \mu=-6.8$, with the
anticipating time $\Delta_{\tau}=|\tau_{2}-\tau_{1}|=3$ . In Fig.
\ref{fig4}(a), We observe that the driven systems anticipates the
driver, at the same time, the amplitude of $x(t+3)$ and $y_{i}$
correlates with each other by $y_{i}(t)=0.5x(t+3)$. Figure
\ref{fig4}(b) shows the time-shifted plot of $x(t+3)$ and $y_{i}$.
It is clear from the scales of the coordinate axes that the slope
of the line is $0.5$. With $\tau_{2}=\tau_{1}$, one can observe
the exact projective synchronization for parameters $\alpha=-1.0,
\beta=6, \tau_{1}=5, \tau_{2}=5, \mu_{1}=-16, \mu_{2}=-0.2,
\mu=16.2$. When $\alpha=-1.0$, we observe the anti-phase
synchronization shown in Fig. \ref{fig5}(a), synchronized chaotic
attractors with an anti-phase pattern where the phase angle
between the synchronized trajectories is $\pi$. From Fig.
\ref{fig5}(b), we see the anti-phase synchronization between
$x(t)$ and $y_{i}(t)$. With $\tau_{2}$ being larger than
$\tau_{1}$, one can observe the projective-lag synchronization for
parameters $\alpha=1.5, \beta=6, \tau_{1}=3, \tau_{2}=5,
\mu_{1}=-16, \mu_{2}=-0.2, \mu=-23.8$. In Fig. \ref{fig6}(a), We
can observe that in this case the response systems lag the state
of the drive system with constant lag time
$\Delta_{\tau}=\tau_{2}-\tau_{1}=2$ and the amplitude of the
driver's and driven's state variables is correlated by
$y_{i}(t)=1.5x(t-2)$. Figure \ref{fig6}(b) shows the time-shifted
plot of $x(t-2)$ and $y_{i}(t)$.

\section{CONCLUSION}

To summarize, we have analytically estimated and numerically
simulated projective-anticipating, projective, and projective-lag
synchronization properties of time-delayed chaotic systems related
to optical bistable or hybrid optical bistable device on random
dynamical networks. Our work may lead to several advantages over
existing work:(1) It is capable of realizing
projective-anticipating and projective-lag synchronization of a
general class of time-delayed chaotic system related to optical
bistable or hybrid optical bistable device on random networks,
while previous work \cite{Hoang1,Hoang} only mentioned
projective-anticipating and projective-lag synchronization on only
two coupled chaotic systems. (2) Although previous work \cite{Hu}
has achieved projective synchronization on complex dynamical
networks, the dynamics of the nodes are coupled partially linear
chaotic systems; we realize projective synchronization of
time-delayed chaotic systems on random dynamical networks, without
the limitation of the partial-linearity; so it can be considered
as an extension of the dynamics of each individual node from
partially linear chaotic systems to non-partially-linear chaotic
systems, or an extension from finite-dimension to
infinite-dimensional chaotic systems. According to the Lyapunov
stability theory, we have achieved the sufficient stability and
necessary conditions for the projective-anticipating, projective,
and projective-lag synchronization manifolds .  We derive that the
transition among projective-anticipating, projective, and
projective-lag synchronizations can be achieved by adjusting the
magnitude relation between the feedback delay time and the
coupling delay time. The validity and feasibility of our method
have been verified by computer simulations of Ikeda and
Mackey-Glass systems.

\begin{acknowledgments}
This work is partially supported by the National Natural Science
Foundation of China (No. 10775060). X.-J.X acknowledges financial
support from FCT (Portugal), Grant No. SFRH/BPD/30425/2006.
\end{acknowledgments}

\bibliographystyle{apsrev}

\end{document}